\documentclass[english,preprint]{revtex4}
\usepackage[T1]{fontenc}
\usepackage[latin9]{inputenc}
\usepackage{textcomp}
\usepackage{relsize}
\usepackage{amsmath}
\usepackage{graphicx}

\makeatletter

\DeclareRobustCommand{\lyxmathsym}[1]{\ifmmode\begingroup\def\b@ld{bold}
  \def\rmorbf##1{\ifx\math@version\b@ld\textbf{##1}\else\textrm{##1}\fi}
  \mathchoice{\hbox{\rmorbf{#1}}}{\hbox{\rmorbf{#1}}}
  {\hbox{\smaller[2]\rmorbf{#1}}}{\hbox{\smaller[3]\rmorbf{#1}}}
  \endgroup\else#1\fi}

\usepackage{hyperref}

\usepackage{latexsym}

\usepackage{xspace}

\@ifundefined{definecolor}
 {\usepackage{color}}{}

\newcommand{\ket}[1]{\ensuremath{|#1\rangle}\xspace}

\makeatother

\usepackage{babel}

\makeatother

\usepackage{babel}

\begin{document}

\title{Single-shot qubit readout in circuit Quantum Electrodynamics}

\author{François $^{1}$Mallet, Florian R. $^{1}$Ong, Agustin $^{1}$Palacios-Laloy,
François $^{1}$Nguyen, Patrice $^{1}$Bertet, Denis $^{1}$Vion$^{\text{*}}$
and Daniel $^{1}$Esteve}

\affiliation{$^{1}$Quantronics group, Service de Physique de l'État Condensé
(CNRS URA 2464), DSM/IRAMIS/SPEC, CEA-Saclay, 91191 Gif-sur-Yvette
Cedex, France}

\email{denis.vion@cea.fr}


\maketitle
\textbf{The future development of quantum information using superconducting
circuits requires Josephson qubits} \cite{Wendin} \textbf{with long
coherence times combined to a high-fidelity readout. Major progress
in the control of coherence has recently been achieved using circuit
quantum electrodynamics (cQED) architectures} \cite{Blais,Wallraff}\textbf{,
where the qubit is embedded in a coplanar waveguide resonator (CPWR)
which both provides a well controlled electromagnetic environment
and serves as qubit readout. In particular a new qubit design, the
transmon, yields reproducibly long coherence times} \cite{transmon_th,transmon_exp}\textbf{.
However, a high-fidelity single-shot readout of the transmon, highly
desirable for running simple quantum algorithms or measuring quantum
correlations in multi-qubit experiments, is still lacking. In this
work, we demonstrate a new transmon circuit where the CPWR is turned
into a sample-and-hold detector, namely a Josephson Bifurcation Amplifer
(JBA)} \cite{siddiqiprl,boaknin}\textbf{, which allows both fast
measurement and single-shot discrimination of the qubit states. We
report Rabi oscillations with a high visibility of $\mathbf{94\%}$
together with dephasing and relaxation times longer than $\mathbf{0.5\,}$\textmu{}s.
By performing two subsequent measurements, we also demonstrate that
this new readout does not induce extra qubit relaxation.}

\begin{figure}[p]
\includegraphics[width=15cm]{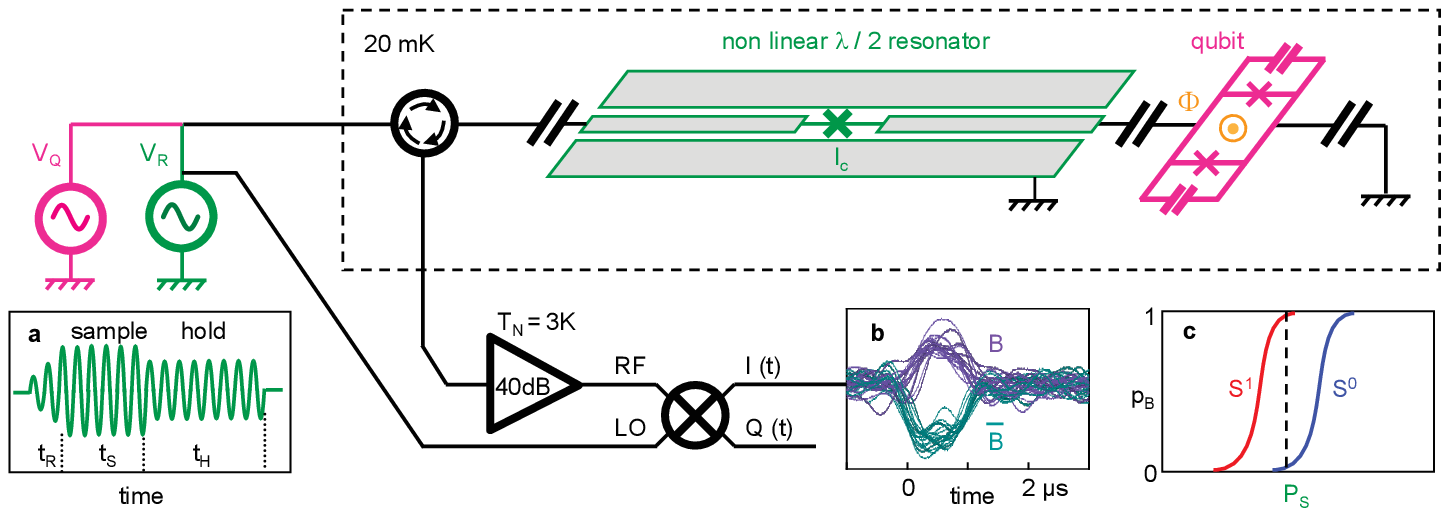}\\

\caption{\textbf{Principle of a single-shot readout for a transmon qubit.}
A transmon (magenta) is capacitively coupled to a coplanar resonator
(green grayed strips) made anharmonic by inserting a Josephson junction
(green cross) at its center. This qubit is coherently driven by a
source $V_{\mathrm{Q}}$ and measured by operating the resonator as
a cavity JBA: a microwave pulse with properly adjusted frequency $f$
and time dependent amplitude (rise, sampling, and holding times $t_{\mathrm{R}}$,
$t_{\mathrm{S}}$, and $t_{H}$, respectively - see inset \textbf{a}
and Methods) is applied by a second source $V_{\mathrm{R}}$; this
pulse is reflected by the system and routed to a cryogenic amplifier
and to a homodyne detection circuit yielding the two quadratures $I$
and $Q$. During the {}``sampling'' time $t_{S}$ the electromagnetic
field in the resonator has a probability $p_{\mathrm{B}}$ to bifurcate
from a low amplitude state $\bar{B}$ to a high amplitude one $B$,
both states corresponding to different amplitudes of $I$ and $Q$.
The {}``holding'' time $t_{H}$ is then used to average $I(t)$
and to determine with certainty if the resonator has bifurcated or
not. \textbf{(b)} Oscillogram showing filtered $I(t)$ traces of both
types (obtained here with $t_{R}=30\,$ns and $t_{S}=t_{H}=250\,$ns).
\textbf{(c)} The probability $p_{\mathrm{B}}$ depends on $f$ and
on the sampling power $P_{\mathrm{S}}$. The two qubit states $\ket{0}$
and $\ket{1}$ shift the resonator frequency, resulting in two displaced
S-curves $S^{0}$ and $S^{1}$. When their separation is large enough,
$P_{S}$ can be chosen (vertical dotted line) so that $\bar{B}$ and
$B$ map $\ket{0}$ and $\ket{1}$ with a high fidelity.}

\label{fig1} 
\end{figure}

A common strategy to readout a qubit consists in coupling it \textit{dispersively}
to a resonator, so that the qubit states $\left|0\right\rangle $
and $\left|1\right\rangle $ shift differently the resonance frequency.
This frequency change can be detected by measuring the phase of a
microwave pulse reflected on (or transmitted through) the resonator.
Such a method, successfully demonstrated with a Cooper pair box capacitively
coupled to a CPWR \cite{Blais,Wallraff}, faces two related difficulties
which have prevented so far from measuring the qubit state in a single
readout pulse (so-called single-shot regime): the readout has to be
completed in a time much shorter than the time $T_{1}$ in which the
qubit relaxes from $\left|1\right\rangle $ to $\left|0\right\rangle $,
and with a power low enough to avoid spurious qubit transitions \cite{bnl-blais}.

This issue can be solved by using a sample-and-hold detector consisting
of a bistable hysteretic system whose two states are brought in correspondence
with the two qubit states. Such a strategy has been implemented in
various qubit readouts \cite{Martinis1,Martinis2}. In our experiment
the bistable system is a Josephson Bifurcation Amplifier (JBA) \cite{siddiqiprl,boaknin}
obtained by inserting a Josephson junction in the middle of the CPWR
(see Fig. 1). When driven by a microwave signal of properly chosen
frequency and power, this non-linear resonator can bifurcate between
two dynamical states $\bar{B}$ and $B$ with different intra-cavity
field amplitudes and reflected phases. In order to exploit the hysteretic
character of this process, we perform the readout in two steps (see
Fig. \ref{fig1}a): the qubit state $\left|0\right\rangle $ or $\left|1\right\rangle $
is first mapped onto $\bar{B}$ or $B$ in a time much shorter than
$T_{1}$; the selected resonator state is then hold by reducing the
measuring power during a time $t_{\mathrm{H}}$ long enough to determine
this stateQ with certainty.

JBAs were used previously to readout quantroniums \cite{siddiqiprb,boulantprb,metcalfeprb}
and flux-qubits, obtaining for the latter fidelities up to 87\% \cite{lupascuprl}
with Quantum-Non-Demolition character \cite{lupascunat}. Here we
couple capacitively a transmon to a JBA, combining all the advantages
of the cQED architecture (long coherence times, scalability) with
the single-shot capability of a sample-and-hold detector. A crucial
characteristic of this new design is its very low back-action during
readout. Indeed the qubit frequency depends only on the slowly-varying
photon number inside the resonator \cite{Schuster}, yielding less
relaxation than in previous experiments where the qubit was coupled
to a rapidly varying variable of the JBA (the intra-resonator current).
Furthermore we designed the resonator to make it bifurcate at a low
photon number, thus avoiding unwanted qubit state transitions during
readout.%
\begin{figure}[p]
\includegraphics[width=15cm]{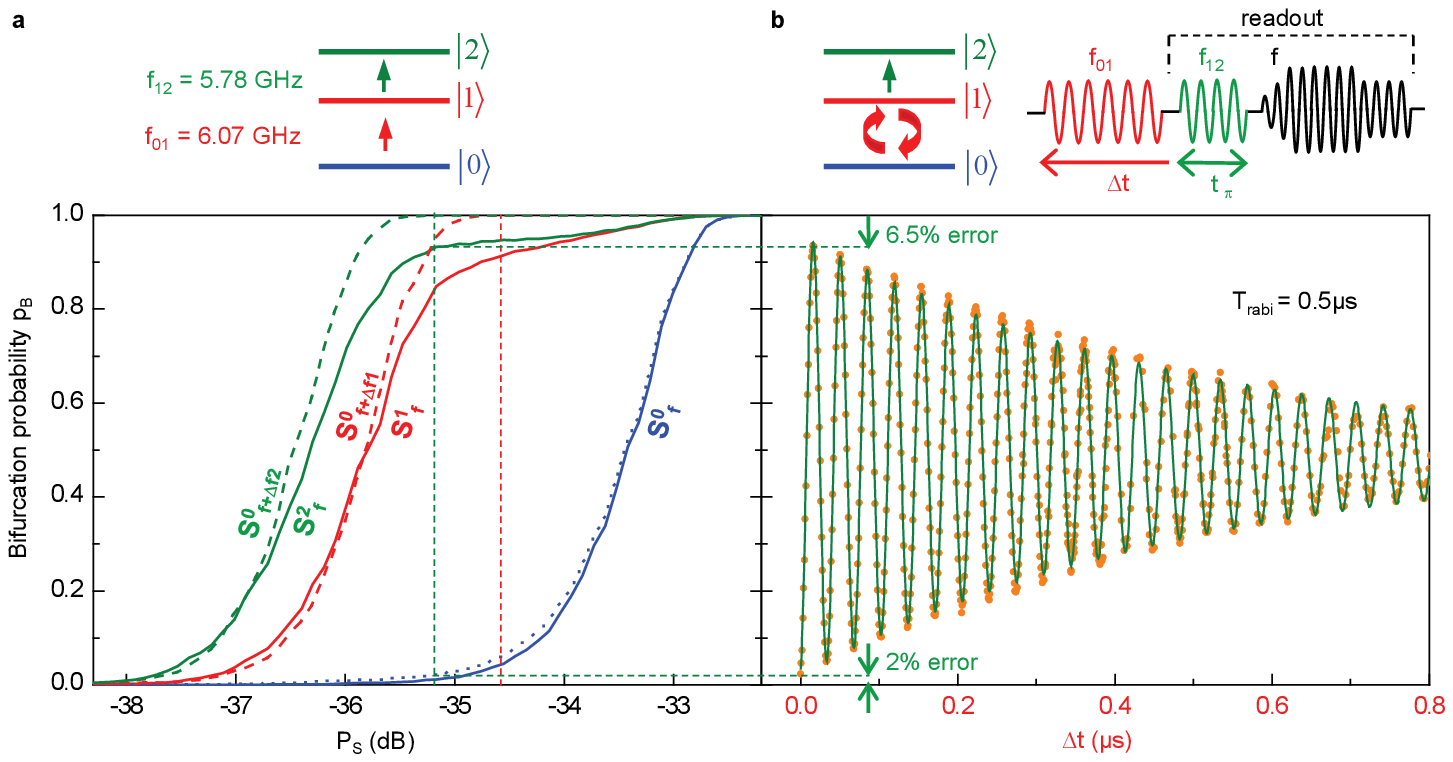}\\

\caption{\textbf{Best single-shot visibity obtained at $\mathbf{\Delta=0.38\,}$GHz
and $\mathbf{f_{\mathrm{C}}-f=17\,}$MHz.} \textbf{(a)} S-curves $p_{\mathrm{B}}(P_{\mathrm{S}})$
obtained with the qubit prepared in state $\ket{0}$, $\ket{1}$,
or $\ket{2}$ (solid lines $S_{\mathrm{f}}^{0}$, $S_{\mathrm{f}}^{1}$
and $S_{\mathrm{f}}^{2}$, respectively) with the proper resonant
$\pi$ pulses (top diagram). The maximum differences between $S_{\mathrm{f}}^{0}$
and $S_{\mathrm{f}}^{1}$ (red vertical line) and between the $S_{\mathrm{f}}^{0}$
and $S_{\mathrm{f}}^{2}$ (green vertical line) define two readout
contrasts of $86\%$ and $92\%$. The readout fidelity is thus increased
by using a composite readout where the measurement pulse is preceded
by a $\pi$ pulse at frequency $f_{12}$ that transfers $\ket{1}$
to $\ket{2}$. The dotted blue curve obtained after a single $\pi$
pulse at frequency $f_{12}$, starting from $\ket{0}$, shows that
this technique has almost no effect on $\ket{0}$. Also plotted are
the curves obtained for $\ket{0}$ when shifting the readout frequency
$f$ by $\Delta{\mathrm{f}}_{1}=4.1\pm0.1\,$ MHz (red dashed line)
and $\Delta{\mathrm{f}}_{2}=5.1\pm0.1\,$ MHz (green dashed line)
in order to match at low $p_{\mathrm{B}}$ the curves obtained for
$\ket{1}$ and $\ket{2}$. The difference between the corresponding
solid and dashed curves is a loss of visibility mostly due to qubit
relaxation before bifurcation. \textbf{(b)} Rabi oscillations at $29\,$
MHz measured with the composite readout, as sketched on top. Dots
are experimental values of $p_{\mathrm{B}}(\Delta t)$ whereas the
solid line is a fit by an exponentially damped sine curve with a $0.5\,$\textmu{}s
decay time and an amplitude of $94\%$ (best visibility). The total
errors in the preparation and readout of the $\ket{0}$ and $\ket{1}$
states are $2\%$ and $6.5\%$ respectively.}

\label{fig2} 
\end{figure}

The complete setup is shown in Fig. \ref{fig1}: the transmon \cite{transmon_th,transmon_exp}
of frequency $f_{01}$ tunable with a magnetic flux $\phi$ is coupled
with a coupling constant $g=44\pm3\,$MHz to the non-linear CPWR of
fundamental frequency $f_{\mathrm{C}}=6.4535\,$GHz, quality factor
$Q_{0}=685\pm15$ and Josephson junction critical current $I_{\mathrm{C}}=0.72\pm0.04\,$\textmu{}A.
In this work the qubit is operated at positive detunings $\Delta=f_{\mathrm{C}}-f_{01}$
larger than $g$. In this dispersive regime the resonator frequency
$f_{\mathrm{Ci}}$ depends on the qubit state $\left|i\right\rangle $,
and the difference $2\chi=f_{\mathrm{C0}}-f_{\mathrm{C1}}$ (so-called
cavity pull) is a decreasing function of $\Delta$. Readout pulses
(Fig. \ref{fig1}a) of frequency $f$ and maximum power $P_{\mathrm{S}}$
are sent to the circuit; after reflection on the resonator their two
quadratures \textit{$I$} and \textit{$Q$} are measured by homodyne
detection. They belong to two clearly resolved families of trajectories
(Fig. \ref{fig1}b) corresponding to both oscillator states $\bar{B}$
and $B$. The escape from $\bar{B}$ to $B$ is a stochastic process
activated by thermal and quantum noise in the resonator \cite{dykman,Vijaythesis},
and occurs during the sampling time $t_{\mathrm{S}}$ with a probability
$p_{\mathrm{B}}$ that increases with $P_{\mathrm{S}}$. The position
of the so-called {}``S-curve'' $p_{\mathrm{B}}(P_{\mathrm{S}})$
depends on the detuning $f_{\mathrm{Ci}}-f$ \cite{siddiqiprl} and
thus on the qubit state. When the two S-curves $S_{\mathrm{f}}^{0}$
and $S_{\mathrm{f}}^{1}$ corresponding to $\ket{0}$ and $\ket{1}$
are sufficiently separated, one can choose a value of $P_{\mathrm{S}}$
at which these states are well mapped onto $\bar{B}$ and $B$ (Fig.
\ref{fig1}c).

We now present our best visibity, obtained at $\Delta=0.38\,$GHz
in this work and confirmed on another sample. We measure $S_{\mathrm{f}}^{0}$
and $S_{\mathrm{f}}^{1}$ (Fig. \ref{fig2}) after preparing the transmon
in state $\ket{0}$ or $\ket{1}$ using a resonant microwave pulse.
The contrast, defined as the maximum difference between both curves,
reaches $86\%$. To interpret the power separation between the S-curves,
we search the readout frequency $f+\Delta{\mathrm{f}}_{1}$ that makes
$S_{{\mathrm{f}}+\Delta{\mathrm{f}}_{1}}^{0}$ coincide with $S_{\mathrm{f}}^{1}$
at low bifurcation probability. This indirect determination of the
cavity pull gives $\Delta{\mathrm{f}}_{1}=4.1\,$MHz, in good agreement
with the value $2\chi=4.35\,$MHz calculated from the experimental
parameters. At high $p_{\mathrm{B}}$ however the two S-curves do
not coincide, which reveals that the limiting factor of our readout
fidelity is relaxation of the qubit before the time needed for the
resonator to reach its final state. To reduce this effect and improve
the readout contrast, we transfer state $\ket{1}$ into the next excited
state $\ket{2}$ with a resonant $\pi$ pulse just before the readout
pulse, yielding the S-curve $S_{\mathrm{f}}^{2}$ and a $92\%$ contrast.
This technique, already used with other Josephson qubits \cite{Martinis2},
is analogous to electron shelving in atomic physics and relies here
on the very low decay rate from $\ket{2}$ to $\ket{0}$ in the transmon.
Figure \ref{fig2}b shows Rabi oscillations between $\ket{0}$ and
$\ket{1}$ obtained with such a composite readout pulse. The visibility,
defined as the fitted amplitude of the oscillations, is $94\%$, and
the Rabi decay time is $0.5\,$\textmu{}s. Of the remaining $6\%$
loss of visibility we estimate that about $4\%$ is due to relaxation
before bifurcation and $2\%$ to residual out-of-equilibrium population
of $\ket{1}$ and to control pulse imperfections. Such a visibility
higher than $90\%$ is in agreement with the width of the S-curves
estimated from numerical simulations, with their theoretical displacement,
and with the measured qubit relaxation time.

\begin{figure}[p]
\includegraphics[width=8cm]{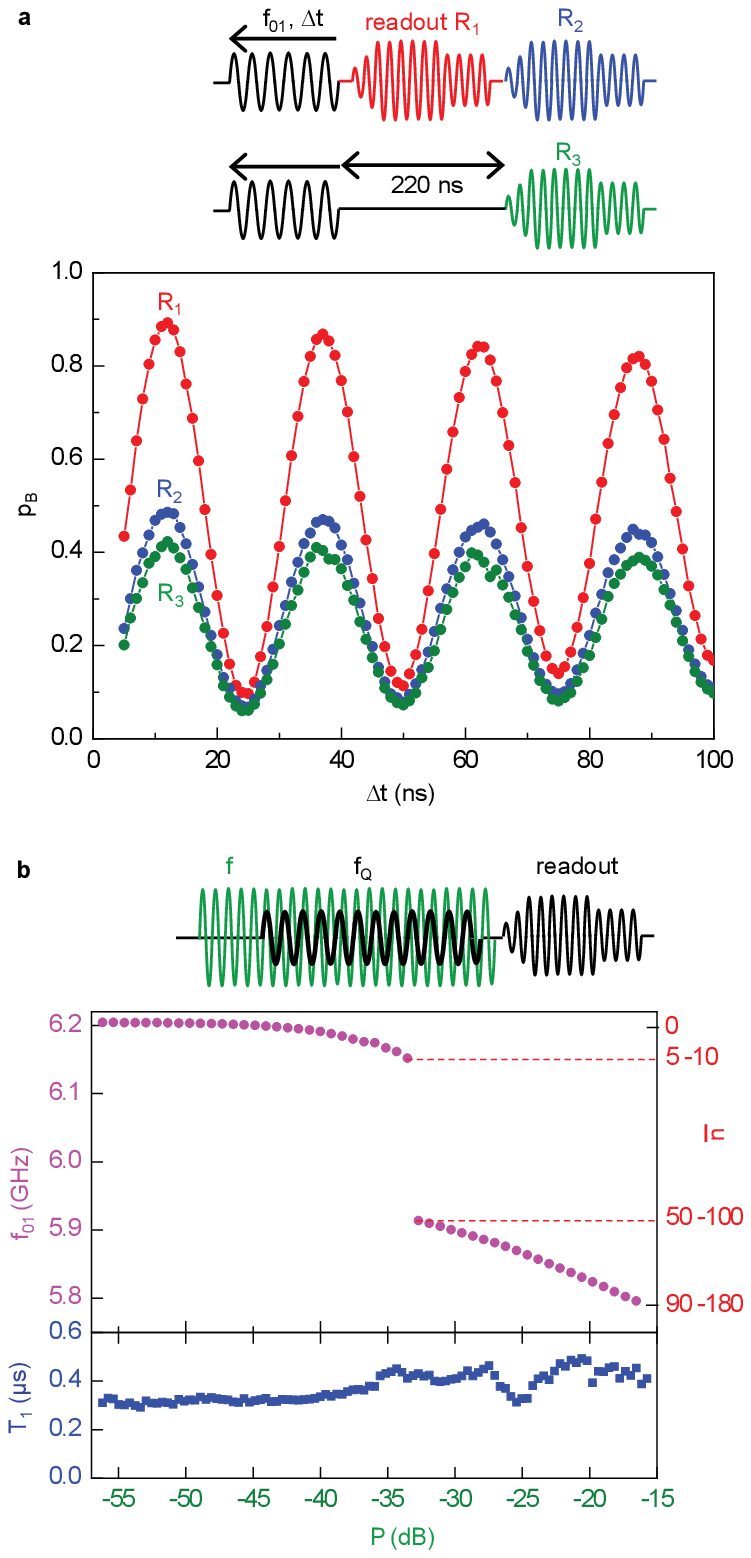}\\

\caption{\textbf{Effect of the readout process onto the qubit at $\mathbf{\Delta=0.25\,}$GHz
and $\mathbf{f_{\mathrm{C}}-f=25\,}$MHz.} \textbf{(a)} Rabi oscillations
$p_{\mathrm{B}}(\Delta t)$ obtained at $P_{\mathrm{S}}=-30.5\,$dB
with the protocols sketched on top, i.e. with two successive readout
pulses placed immediately after the control Rabi pulse (red and blue
dots), or with the second pulse only (green dots). The loss of Rabi
visibility between the red curve ($83\%$) and the blue ($44\%$)
and green ($37\%$) ones is due to qubit relaxation during the first
readout or the delay. \textbf{(b)} Top panel: Spectroscopic determination
of the qubit frequency $f_{01}$ when it is AC-Stark shifted by an
auxiliary microwave with frequency $f$ and power $P$ (protocol on
top). The shift provides an in-situ estimate of the average photon
number $\bar{n}$ in the resonator (right scale) with a precision
of $\pm30\%$. The bifurcation is seen as a sudden jump. Bottom panel:
qubit relaxation time $T_{1}$ (measurement protocol not shown) in
presence of the same auxiliary field. $T_{1}$ does not show any strong
decrease even at power well above bifurcation.}

\label{fig3} 
\end{figure}

The visibility being limited by relaxation, it is important to determine
whether the readout process itself increases the qubit relaxation
rate. For that purpose we compare (at $\Delta=0.25\,$GHz) Rabi oscillations
obtained with two different protocols: the control pulse is either
followed by two successive readout pulses yielding curves R$_{1}$
and R$_{2}$, or by only the second readout pulse yielding curve R$_{3}$
(see Fig. \ref{fig3}a). R$_{2}$ and R$_{3}$ exhibit almost the
same loss of visibility compared to R$_{1}$, indicating that relaxation
in the presence of the first readout pulse is the same as (and even
slightly lower than) in its absence.

To further investigate this remarkable effect, we measure $T_{1}$
in presence of a microwave field at the same frequency $f$ as during
readout, and for different input powers $P$ (see Fig. \ref{fig3}b).
We first roughly estimate the intra-cavity mean photon number $\bar{n}(P)$
by measuring the AC-Stark shifted qubit frequency $f_{01}(P)$ \cite{Schuster}
(the correpondence $f_{01}(n)$ is obtained by a numerical diagonalization
of the Hamiltonian of the transmon coupled to a field mode with $n$
photons). Bifurcation is clearly revealed by a sudden jump of $\bar{n}$
from about 5-10 to 50-100 photons. Meanwhile $T_{1}$ does not show
any decrease up to about $5\,$dB above bifurcation. It even slightly
increases because the qubit frequency is pushed away from the cavity,
slowing down spontaneous emission as explained in the next paragraph.
This is in strong contrast with all previous experiments using a JBA
readout \cite{picot,Vijaythesis}. These results prove that our design
achieves very low back-action on the qubit. A similar behavior was
observed for most qubit frequencies, except at certain values of $P$
and $f_{01}$ where dips in $T_{1}(P)$ were occasionally observed
above bifurcation.

\begin{figure}[p]
\includegraphics[width=8cm]{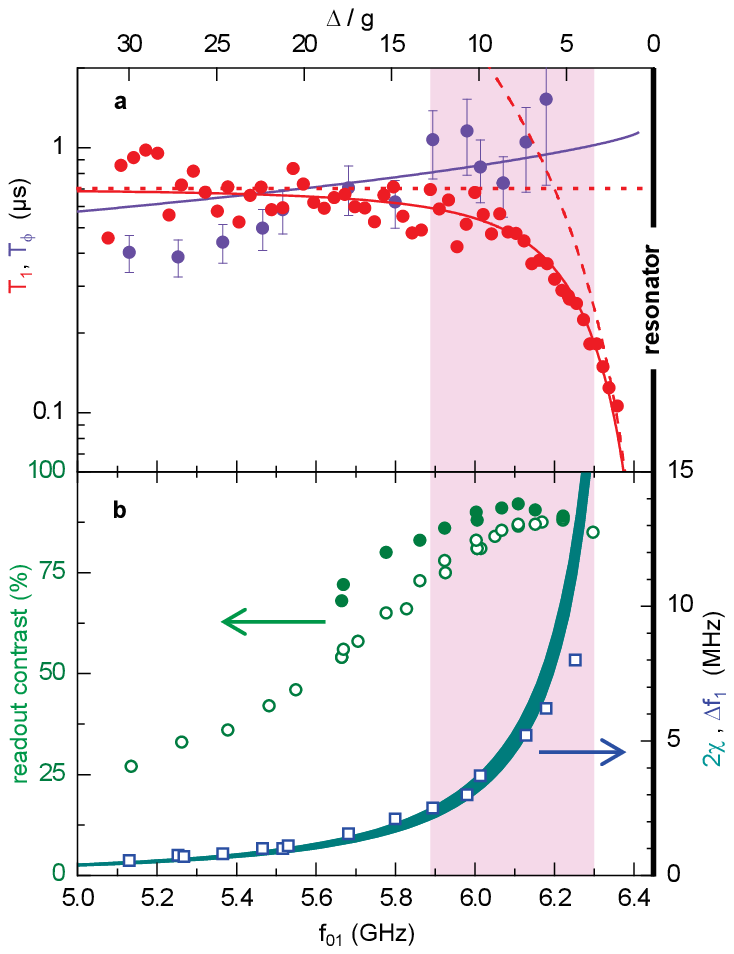}\\

\caption{\textbf{Trade-off between qubit coherence and readout fidelity.} \textbf{(a)}
Experimental relaxation time $T_{1}$ (red dots) and dephasing time
$T_{\phi}$ (violet dots) of the qubit as a function of $f_{01}$
(or equivalently $\Delta/g$). Note that $T_{\mathrm{\phi}}\approx2.5\pm0.5\,$\textmu{}s
at the flux optimal point \cite{quantronium} ($\Delta\approx$ -0.75
GHz, data not shown). Error bars on $T_{\mathrm{\phi}}$ are absolute
minima and maxima resulting from the maximum experimental uncertainties
on the coherence times $T_{1}$ and $T_{2}$ (see methods). The solid
red line is the value of $T_{1}$ obtained by adding to the expected
spontaneous emission through the resonator (dashed red line) a relaxation
channel of unknown origin with $T_{1}=0.7\,$\textmu{}s (horizontal
dotted line). The blue line is the pure dephasing time $T_{\mathrm{\phi}}$
corresponding to a $1/f$ flux noise with an amplitude set to $20\,\lyxmathsym{\textmu}{\mathrm{\phi}}_{0}/\sqrt{\mathrm{Hz}}$
at $1\,$Hz. (\textbf{b} - left scale) Readout contrast with (green
dots) and without (green circles) transfer from state $\ket{1}$ to
$\ket{2}$ (see Fig. $2$). (\textbf{b} - right scale) Effective cavity
pull $\Delta{\mathrm{f}}_{1}$ (blue squares) determined as shown
in Fig. \ref{fig2}. For the sake of comparison, the predicted cavity
pull $2\chi$ in the dispersive approximation is also shown as a cyan
region, taking into account the maximal experimental uncertainty on
$g$. The pink area denotes the region where the readout contrast
is higher than $85\%$.}

\label{fig4} 
\end{figure}

We now discuss the dependence of the readout contrast and qubit coherence
on the detuning $\Delta$. Besides acting as a qubit state detector,
the resonator serves also as a filter protecting the qubit against
spontaneous emission into the 50 $\mathrm{\Omega}$ impedance of the
external circuit \cite{esteve,transmon_exp2}. The smaller $\Delta$,
the stronger the coupling between the qubit and the resonator, implying
a larger separation between the $S_{\mathrm{f}}^{0}$ and $S_{\mathrm{f}}^{1}$
curves but also a faster relaxation. We thus expect the contrast to
be limited by relaxation at small $\Delta$, by the poor separation
between the S-curves at large $\Delta$, and to exhibit a maximum
in between. Figure \ref{fig4} presents a summary of our measurements
of contrast and coherence times. At small $\Delta$, $T_{1}$ is in
quantitative agreement with calculations of the spontaneous emission
trough the resonator. However it presents a saturation, similarly
as observed in previous experiments \cite{transmon_exp2}, but at
a smaller value around $0.7\,$\textmu{}s. The effective cavity
pull $\Delta{\mathrm{f}}_{1}$ determined from the S-curves shifts
(cf. Fig. \ref{fig2}) is in quantitative agreement with the value
of $2\chi$ calculated from the sample parameters. The contrast varies
with $\Delta$ as anticipated and shows a maximum of $92\%$ at $\Delta=0.38\,$
GHz, where $T_{1}=0.5\,$\textmu{}s. Larger $T_{1}$ can be obtained
at the expense of a lower contrast and reciprocally. Another important
figure of merit is the pure dephasing time $T_{\mathrm{\phi}}$ \cite{ithier}
which controls the lifetime of a superposition of qubit states. $T_{\mathrm{\phi}}$
is extracted from Ramsey fringes experiments (see Methods), and shows
a smooth dependence on the qubit frequency, in qualitative agreement
with the dephasing time deduced from a $1/f$ flux noise of spectral
density set to $20\,$\textmu{}${\mathrm{\phi}}_{0}/\sqrt{\mathrm{Hz}}$
at $1\,$Hz, a value similar to those reported elsewhere \cite{wellstood}.
To summarize our circuit performances, we obtained a $400\,$MHz frequency
range (pink area on Fig. \ref{fig4}) where the readout contrast is
higher than $85\%$, $T_{1}$ is between $0.7\,$\textmu{}s and
$0.3\,$\textmu{}s, and $T_{\phi}$ between $0.7\,$\textmu{}s
and $1.5\,$\textmu{}s. Further optimization of the JBA parameters
$I_{\mathrm{C}}$ and $Q_{0}$ could increase this high-visibility
readout frequency window.

In conclusion we have demonstrated the high-fidelity single-shot readout
of a transmon qubit in a circuit-QED architecture using a bifurcation
amplifier. This readout does not induce extra qubit relaxation and
preserves the good coherence properties of the transmon. The high
fidelity achieved should allow a test of Bell's inequalities using
two coupled transmons, each one with its own JBA single-shot readout.
Moreover, our method could be used in a scalable quantum processor
architecture, in which several transmon-JBAs with staggered frequencies
are read by frequency multiplexing.

\section*{Methods}

\subsection{Sample fabrication}

The sample was fabricated using standard lithography techniques. In
a first step, a $120\,$nm-thick niobium film is sputtered on an oxidized
high-resistivity silicon chip. It is patterned by optical lithography
and reactive ion etching of the niobium to form the coplanar waveguide
resonator. The transmon and the Josephson junction of the CJBA are
then patterned by e-beam lithography and double-angle evaporation
of two aluminum thin-films, the first one being oxidized to form the
junction tunnel barrier. The chip is glued on and wire-bonded to a
microwave printed-circuit board enclosed in a copper box, which is
thermally anchored to the mixing chamber of a dilution refrigerator
at typically $20\,$mK.

\subsection{Electrical lines and signals}

Qubit control and readout microwave pulses are generated by mixing
the output of a microwave source with {}``DC'' pulses generated
by arbitrary waveform generators, using DC coupled mixers. They are
then sent to the input microwave line that includes bandpass filters
and attenuators at various temperatures. The powers given in dB in
this letter are arbitrarily refered to $1\,$mW (on $50\,\Omega$)
at the input of the dilution refrigerator; the total attenuation down
to the sample is about $-77\,$dB. The pulses are routed to the resonator
through a circulator to separate the input and output waves.

The readout output line includes a bandpass filter ($4-8\,$GHz),
2 isolators, and a cryogenic amplifier (CITCRYO 1-12 from California
Institute for Technology) with $38\,$dB gain and noise temperature
$T_{\mathrm{N}}=3\,$K. The output signal is further amplified at
room-temperature with a total gain of $56\,$dB, and finally mixed
down using an I/Q mixer with a synchronized local oscillator at the
same frequency. The $I$ and $Q$ quadratures are further amplified
by $20\,$dB, and sampled by a fast digitizer. The data are then transferred
to a computer and processed. The single-shot traces of Fig. \ref{fig1}b.
were obtained with an additional $10\,$ MHz low-pass filter.

\subsection{Sample characterization}

The characteristic energies of the system, namely the transmon Josephson
energy $E_{\mathrm{J}}=21\,$GHz and charging energy $E_{c}=1.2\,$GHz
(for a Cooper-pair), as well as the qubit-resonator coupling constant
$g\,$, have been determined by spectroscopic measurements. The bare
resonator frequency $f_{C}$ was determined at a magnetic field such
that the qubit was far detuned from the resonator.

\subsection{Qubit state preparation}

We prepare the qubit in its ground state with a high fidelity at the
beginning of each experimental sequence by letting it relax during
about $20$\textmu{}s. We estimate at about $1\%$ the equilibrium
population in state $\ket{1}$ due to residual noise coming from measurement
lines.

To prepare the qubit in its excited state $\ket{1}$ or $\ket{2}$,
one or two successive resonant square-shaped pulses of length $t_{\pi}\sim$
20 ns are applied prior to the readout pulse. The dotted blue S-curve
of Fig. \ref{fig1} was recorded with a single resonant $\pi$ pulse
at $f_{12}$ (see text): it reveals that this pulse induces a spurious
population of the $\ket{1}$ state of order $1\%$. We checked that
this effect is corrected by using gaussian-shaped pulses \cite{Martinis1}
(data not shown).

\subsection{Readout Pulses}

We give here more information on the timing of the readout pulses
used is this work. In Fig. \ref{fig2}, readout is performed at \textbf{$f_{\mathrm{C}}-f=17\,$}MHz,
and we used $t_{R}=15\,$ns, $t_{\mathrm{S}}=250\,$ns and $t_{\mathrm{H}}=700\,$ns.
We stress that although $t_{\mathrm{S}}$ is of the same order of
magnitude as $T_{1}$, the observed relaxation-induced loss of contrast
is rather low, which may seem surprising. This is due to an interesting
property of our readout : when the qubit is in state $\ket{1}$, the
JBA bifurcates with a high probability, implying that all bifurcation
events occur at the very beginning of the readout pulse (instead of
being distributed exponentially during $t_{S}$). We nevertheless
keep $t_{\mathrm{S}}=250\,$ns because the bifurcation process itself
needs such a duration to develop properly. The effective measurement
time $t_{M}$ is thus shorter than $t_{S}$. We verified that weighted
sums of $S_{\mathrm{f}}^{0}$ and $S_{{\mathrm{f}}+\Delta{\mathrm{f}}_{i}}^{0}$
fit properly the $S_{\mathrm{f}}^{i}$ curves (i=1,2) of Fig. \ref{fig2},
allowing us to quantify the population of each level at readout. Using
the experimentally determined relaxation times $T_{1}^{2\rightarrow1}\sim0.3\,$\textmu{}s
and $T_{1}^{1\rightarrow0}\sim0.45\,$\textmu{}s, we thus estimate
$t_{M}\sim40\,$ns.

In Fig. \ref{fig3}, readout is performed at \textbf{$f_{\mathrm{C}}-f=25\,$}MHz,
to reduce the total measurement duration. Indeed, as a larger readout
detuning implies a higher driving power and thus a higher reflected
power, the signal to noise ratio is increased which allows to shorten
$t_{\mathrm{H}}\,$ to $50\,$ns. We also used for these data $t_{\mathrm{R}}=10\,$ns
and $t_{\mathrm{S}}=40\,$ns to shorten the overall measurement time,
which also decreases the maximal contrast to approx $83\%$. Finally,
a delay time of $120\,$ ns between the two readout pulses has been
optimized experimentally to empty the resonator of all photons due
to the first measurement, and thus avoid any spurious correlations
between the two outcomes of the sequence.

\subsection{Coherence times measurement}

The qubit coherence times are measured using standard experimental
sequences \cite{quantronium}. For the relaxation time $T_{1}$, we
apply a $\pi$ pulse and measure the qubit state after a variable
delay, yielding an exponentially decaying curve whose time constant
is $T_{1}$. The coherence time $T_{2}$ is obtained by a Ramsey experiment:
two $\pi/2$ pulses are applied at a frequency slightly off-resonance
with the qubit and with a variable delay; this yields an exponentially
damped oscillation whose time constant is $T_{2}$. We then extract
the pure dephasing contribution $T_{\mathrm{\phi}}$ to the quantum
coherence (as well as the associated maximum uncertainty) using the
relation $T_{\mathrm{\phi}}^{-1}=T_{\mathrm{2}}^{-1}-(2T_{1})^{-1}$
\cite{ithier}.
\begin{acknowledgments}
We acknowlege financial support from European projects EuroSQIP and
Midas, from ANR-08-BLAN-0074-01, and from Region Ile-de-France for
the nanofabrication facility at SPEC. We gratefully thank P. Senat
and P. Orfila for technical support, and acknowledge useful discussions
within the Quantronics group and with A. Lupascu, I. Siddiqi, M. Devoret
and A. Blais.

\bigskip{}

Author contributions: F.M., P.B., D.V. \& D.E. designed the experiment,
F.O. fabricated the sample, F.M., F.N., A.P.L, F.O. \& P.B. performed
the measurements, and all the authors contributed to the writing of
the manuscript.

\bigskip{}

Correspondance should be addressed to D.V.\end{acknowledgments}

\end{document}